\documentclass{pramana}
\usepackage{graphicx,amsmath,bm}
\usepackage[usenames]{color}
\usepackage{graphicx}
\usepackage{amssymb}
\usepackage{amsmath}
\DeclareMathOperator{\sech}{sech}
\usepackage{braket}
\usepackage{epstopdf}
\usepackage{placeins}
\usepackage{bm}
\usepackage{hyperref}
\usepackage[normalem]{ulem}
\usepackage[utf8]{inputenc}
\usepackage{float}
\usepackage{ragged2e} 
\usepackage{balance} 
\usepackage{caption}
\usepackage{lastpage}
\usepackage[numbers,sort&compress]{natbib}
\begin{document}
	
	%% paper title
	\title{PINN-Based Framework for Soliton Solutions of Gross--Pitaevskii and Nonlinear Schr\"odinger Equations}
	
	%% author names
	%% use \textsuperscript for affiliations
	%% use * for corresponding author
	
	\author{
		P.~S.~Vinayagam\textsuperscript{1,*},
		Sai Jeevanth G\textsuperscript{2},
		D.~Aravindha~Krishnan\textsuperscript{3}
		\and
		Nithish Kathiravan\textsuperscript{4}
	}
	
	\affilOne{\textsuperscript{1}
		Department of Physics, PSG College of Arts and Science, Avinashi Road,
		Civil Aerodrome Post, Coimbatore--641014, Tamil Nadu, India\par}
	
	\affilTwo{\textsuperscript{2}
		Department of Physics, Vellore Institute of Technology,
		Vellore--632014, Tamil Nadu, India\par}
	
	\affilThree{\textsuperscript{3}
		Department of Physics, Pondicherry University,
		Kalapet, Puducherry--605014, India\par}
	
	\affilFour{\textsuperscript{4}
		Department of Biotechnology, PSG College of Arts and Science, Avinashi Road,
		Civil Aerodrome Post, Coimbatore--641014, Tamil Nadu, India\par}
	
	%% escape two-column mode for title, affiliations, abstract
	\twocolumn[{
		
		\maketitle
		
		%% corresponding author email
		\corres{psvinayagam11@gmail.com}
		
		%% manuscript information
		\msinfo{Received date}{Revised date}{Accepted date}
		
		%% abstract
		\begin{abstract}
			This study presents a data-driven framework for solving nonlinear wave equations,
			specifically the Gross--Pitaevskii equation (GPE) and the single-component nonlinear
			Schr\"odinger equation (NLSE), using physics-informed neural networks (PINNs).
			The approach integrates physical constraints directly into the neural network loss
			function, enabling efficient training without labeled data. We implement a PINN-based
			framework for solitons that models a variety of localized wave structures across both
			equations. Predicted solutions are compared with exact analytical results and show strong
			agreement with low error. The method effectively captures soliton profiles in both the
			GPE and NLSE. The accuracy and flexibility of the framework suggest its usefulness for
			studying nonlinear differential equations relevant to Bose--Einstein condensates and
			nonlinear optics.
		\end{abstract}
		
		%% keywords
		\keywords{Solitons, Breathers, Physics-informed neural networks, Darboux transformation, Deep learning}
		
		%% PACS numbers
		\pacs{05.45.Yv; 03.75.Lm; 42.65.Tg; 07.05.Mh}
		
	}]
	
	\section{Introduction}
	
	Understanding nature begins with the study of physical systems that evolve over time. Any change occurring in such a system is typically interpreted in relation to time. This study of time-dependent changes is referred to as the analysis of dynamical systems. These systems are generally described using differential equations, which can be categorized as either linear or nonlinear based on the system's behavior over time~\cite{lakshmanan2003nonlinear}.
	
	If the system's behavior repeats consistently over fixed intervals, it is described as linear. Conversely, nonlinear systems exhibit complex, irregular responses that do not follow a fixed pattern. The mathematical formulation of both types aligns with linear and nonlinear differential equations, respectively.
	
	Among these, the Gross-Pitaevskii Equation (GPE) and the Nonlinear Schrödinger Equation (NLSE) are two prominent nonlinear differential equations. The GPE models mean-field interactions in Bose-Einstein Condensates (BECs)~\cite{dalfovo1999theory}, while the NLSE describes the propagation of optical solitons in fibers. Physicists typically solve these equations either analytically—to obtain exact soliton solutions—or numerically to approximate them~\cite{radha2015,bao2003numerical}.
	
	Analytically, techniques such as the gauge transformation and Darboux transformation are widely used for generating systematic, multi-order soliton solutions. Numerically, methods like the Runge-Kutta algorithm, Split-step Crank-Nicolson, and Fourier-based techniques are commonly employed. Analytical methods, though powerful, are increasingly reliant on symbolic computation due to their complexity. Numerical methods, while more accessible, can suffer from inefficiencies—especially when applied to higher-dimensional problems or complex boundary conditions.
	
	In recent years, deep learning approaches~\cite{zhou2022}, particularly Physics-Informed Neural Networks (PINNs)~\cite{tian2023}, have emerged as promising tools for solving differential equations~\cite{chen2024}. PINNs combine the universal approximation capabilities of neural networks with embedded physical laws, encoded in the loss function using automatic differentiation~\cite{pu2023}. By treating governing equations as soft constraints, PINNs can produce highly accurate solutions without requiring large labeled datasets—especially advantageous for problems in high-dimensional spaces or where experimental data is limited.
	
	Unlike traditional numerical solvers, PINNs can manage irregular geometries and learn dynamics in both spatial and temporal domains, making them particularly effective for modeling nonlinear wave phenomena in GPE and NLSE systems.
	
	In this study, we investigate the application of PINNs to solving nonlinear equations such as the GPE and NLSE. The governing equations are embedded directly into the loss function, enabling the neural network to approximate solutions without relying on labeled training data. We focus on validating the accuracy of this method by comparing PINN-generated solutions against established analytical and numerical results. Furthermore, we evaluate its performance across varying nonlinear and potential strengths, demonstrating the model's effectiveness in solving complex quantum systems with high precision~\cite{wen2022}. The accuracy and efficiency of deep learning method is handling complex nonlinear systems is established by the structural comparison of solutions obtained by PINN and conventional methods~\cite{li2025}.
	
	The paper is presented as follows: We introduce the mathematical formulation of the GPE and NLSE, highlighting their solitonic solutions in Section~\ref{sec:pinn}. Section~\ref{sec:framework}, dealing the PINN methodology in detail and the process of performing it for the model is discussed. In section~\ref{sec:data} we discussed our results obtained by PINN method and Finally, Section~\ref{sec:summary} concludes the paper with a summary of our findings and potential directions for future research. 
	
	\section{Physics-Informed Neural Networks Framework}
	\label{sec:pinn}
	
	Physics-Informed Neural Networks (PINNs) integrate physical laws, expressed as partial differential equations (PDEs), directly into the training process via the loss function, eliminating the need for labeled datasets~\cite{raissi2019physics}. This makes them highly data-efficient and capable of solving complex physical systems with minimal supervision.
	
	A major strength of PINNs lies in bypassing mesh-based discretization, which often limits traditional numerical methods. This enables more flexible problem modeling, particularly in systems with intricate boundary conditions or multi-scale dynamics such as quantum fluids, Bose-Einstein condensates, and nonlinear optical media~\cite{wen2022}. Compared to classical solvers, PINNs offer improved scalability and reduced computational overhead in high-dimensional simulations~\cite{song2024}. The framework's core advantage is its generalizability. Once trained, a PINN model can adapt to varying parameters and conditions without re-solving from scratch~\cite{tian2023,sun2024}. This is especially beneficial for analyzing solitonic dynamics under different interaction regimes or potentials~\cite{pu2023}. Furthermore, PINNs support integration with experimental data, offering a bridge between theoretical modeling and experimental validation in domains like ultracold atoms and optical systems~\cite{zhou2022}.
	
	By leveraging automatic differentiation and deep learning optimizers, PINNs maintain physical consistency across initial and boundary conditions, making them a powerful alternative to conventional solvers for studying nonlinear wave propagation~\cite{zhang2025}. Given a PDE of the form
	\begin{equation}
		\mathcal{F}(u, u_x, u_t, u_{xx}, u_{tt}, \dots) = 0, \quad (x, t) \in \Omega \times [0,T],
	\end{equation}
	where $\mathcal{F}$ is a differential operator, PINNs approximate the solution $u(x, t)$ using a neural network $u_{\theta}(x, t)$, parameterized by $\theta$. The network is trained by minimizing a loss function that consists of three terms:
	\begin{equation}
		\mathcal{L} = \mathcal{L}_{\text{PDE}} + \lambda_b \mathcal{L}_{\text{BC}} + \lambda_i \mathcal{L}_{\text{IC}},
	\end{equation}
	where $\mathcal{L}_{\text{PDE}}$ enforces the PDE residual by penalizing deviations from the governing equation,
	\begin{equation}
		\mathcal{L}_{\text{PDE}} = \sum_{i=1}^{N_r} \left| \mathcal{F}(u_{\theta}, u_{\theta, x}, u_{\theta, t}, u_{\theta, xx}, u_{\theta, tt}, \dots) \right|^2,
	\end{equation}
	$\mathcal{L}_{\text{BC}}$ ensures adherence to boundary conditions, and $\mathcal{L}_{\text{IC}}$ enforces the initial conditions. Training PINNs involves optimizing the parameters $\theta$ using gradient-based methods such as Adam and L-BFGS. PINNs have been successfully applied to solving nonlinear Schrödinger equations (NLSE), Burgers' equation, and the Gross-Pitaevskii equation (GPE)~\cite{rao2022,yeo2023}, demonstrating their capability in high-dimensional and complex dynamical systems~\cite{raissi2019physics}.
	
	By embedding the governing equations, boundary conditions, and initial conditions into the training process, PINNs provide a flexible framework for solving PDEs.
	
	The key idea behind Physics-Informed Neural Networks (PINNs) is to approximate the solution $u(x, t)$ of a partial differential equation (PDE) using a neural network $u_\theta(x, t)$, where $\theta$ represents the trainable parameters. The loss function for training consists of multiple components. The PDE residual loss measures the deviation of $u_\theta$ from satisfying the governing equation. The boundary condition loss ensures that the solution adheres to the specified boundary conditions, while the initial condition loss enforces consistency with the given initial conditions.
	
\begin{figure*}[!htbp]
		\centering
		\includegraphics[width=0.9\textwidth]{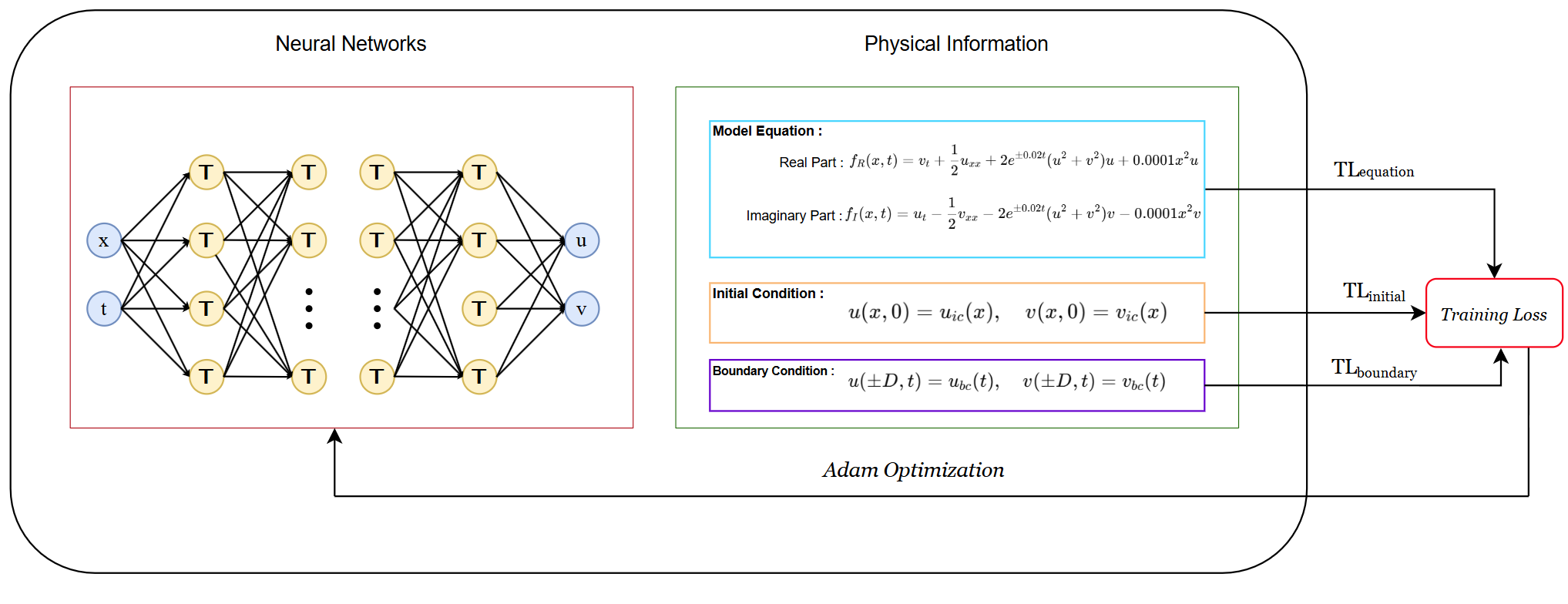}
		\caption{Architecture of the Physics-Informed Neural Network (PINN) designed to solve the generalized Gross-Pitaevskii equation (GPE). The neural network maps spatio-temporal coordinates $(x, t)$ to the real $(u)$ and imaginary $(v)$ components of the solution.}
		\label{fig:pinn_gpe}
	\end{figure*}

	In Figures~\ref{fig:pinn_nlse} and~\ref{fig:pinn_gpe}, we have shown the intended framework of the proposed Physics-Informed Neural Network (PINN). The framework has been designed into LEFT, MIDDLE, and RIGHT. The neural scheme on the left side admits the inputs like spatial ($x$) and temporal ($t$) coordinates. The received inputs travel through several hidden layers made up of neurons with trainable parameters in the middle. Finally, we have the real ($u$) and imaginary ($v$) components of the wave function as output on the right side. The system's physical restrictions are encoded in the right section. The network learns a physically consistent solution by dividing the NLSE into real and imaginary parts and integrating nonlinear interactions through the term $(|u|^2 + |v|^2)$. Three parts make up the loss function: one enforces the NLSE itself, another makes sure that beginning conditions are followed, and a third imposes boundary restrictions. The Adam optimization approach is used to reduce the total training loss, which is mostly caused by these loss components.
	
	In order to direct the network toward precise solutions, the training loss ($TL$) incorporates constraints from the NLSE, initial, and boundary conditions, as illustrated in both Figures~\ref{fig:pinn_nlse} and~\ref{fig:pinn_gpe}. Consistency with the NLSE and GPE is ensured via the implicit differentiation technique that the PINN framework offers to impose the governing equation at collocation locations. PINNs are especially well-suited for solving differential equations where conventional numerical methods encounter difficulties due to the integration of physical laws into the learning process~\cite{karniadakis2021physics}. This method is useful for complex nonlinear partial differential equations such as NLSE and GPE etc, related problems in quantum physics, optics, and Bose-Einstein condensates where nonlinear interactions control the evolution of the wave function. The images illustrates how PINNs, by utilizing deep learning approaches, are an effective tool for getting approximation solutions to complex dynamical systems. 
	
	The physical information module fulfils the compliance of the GPE requirements in the neural network in the figure. The incorporation of the model equation is done by the residual terms $f_R(x,t)$ and $f_I(x,t)$ as real and imaginary components associated with the model. Boundary conditions limit the function behavior at $x = \pm D$, and the initial condition defines the wavefunction at time $t=0$. The total loss function comprises three components: $TL_{\text{equation}}$, $TL_{\text{initial}}$, and $TL_{\text{boundary}}$. Using the Adam Optimization approach, we optimize the weights iteratively to minimize the overall training loss.
	
	This integration of physical knowledge enables PINNs to solve PDEs without requiring labeled datasets, making them highly suitable for problems like the NLSE.
	
	\section{PINN Framework for GPE and NLSE}
	\label{sec:framework}
	
	In this section, we present the implementation and configuration details of our PINN Framework for Solitons, developed to solve nonlinear partial differential equations such as the Gross–Pitaevskii Equation (GPE) and the Nonlinear Schrödinger Equation (NLSE). 
	
	To ensure reproducibility of the PINN Framework for Solitons, we provide a comprehensive description of the implementation details used for solving both the Gross–Pitaevskii Equation (GPE) and the Nonlinear Schrödinger Equation (NLSE).
	
	\subsection{Gross–Pitaevskii equation (GPE)}
	
	The three-dimensional mean field equation of motion, which describes the evolution of its macroscopic wave function, is known as the Gross–Pitaevskii equation (GPE). Studying this GPE, helps us to understand the dynamics and collective behaviour of Bose–Einstein condensate~\cite{gross1961structure}.
	\begin{equation}\label{eq:general_gpe}
		i\hbar \frac{\partial \psi(\vec{r}, t)}{\partial t} = 
		\left[ -\frac{\hbar^2 \nabla^2}{2m} + V_{\text{ext}}(\vec{r}) + g |\psi(\vec{r}, t)|^2 \right] \psi(\vec{r}, t),
	\end{equation}
	where $\psi(\vec{r}, t)$ is the macroscopic wave function, $V_{\text{ext}}$ represents the external harmonic trapping potential, and the interatomic interaction strength is given by $g = \frac{4\pi \hbar^2 a_s}{m}$, with s-wave scattering length $a_s$.
	
	In our case, we consider a system with time-dependent scattering lengths and a parabolic trapping potential, and the modified Gross-Pitaevskii equation~\cite{perez1998} in one dimension is given in the following form.
	\begin{equation}\label{eq:integrable_gpe}
		i \frac{\partial \psi(x,t)}{\partial t} +\frac{1}{2} \frac{\partial^2 \psi(x,t)}{\partial x^2} + 2a(t) |\psi(x,t)|^2 \psi(x,t) 
		+ \frac{1}{4} \zeta^2 x^2 \psi(x,t) = 0
	\end{equation}
	The time-dependent scattering length is given by $a(t)=a_{0}\exp[\zeta t]$, where $a_{0}$ is the scattering length at $t=0$. Setting $a_{0}=1$ and $\zeta=\pm 0.02$, we substitute these values into the equation~(\ref{eq:integrable_gpe}) to obtain a new model equation, which we then analyze using the PINN Framework.
	\begin{equation}\label{eq:unfs_gpe}
		\begin{aligned}
			i\,\psi_t(x,t)
			&\;+\; \frac{1}{2}\,\psi_{xx}(x,t)  \\
			&\;+\; 2 e^{\pm 0.02 t}\, |\psi(x,t)|^{2}\,\psi(x,t)  \\
			&\;+\; 0.0001\, x^{2}\, \psi(x,t)
			= 0.
		\end{aligned}
	\end{equation}

	The inital condition $\psi_{ic}$, lower boundary condition $\psi_{lb}$ and upper boundary condition $\psi_{ub}$ in the domain $D$ under consideration for the above model equation~(\ref{eq:unfs_gpe}) are given as
	\begin{subequations}\label{eq:conditions_gpe}
		\begin{align}
			\psi_{ic}(x)&=\psi(x,0), \quad x\in[-D,D],\\
			\psi_{lb}(t)&=\psi(-D,t), \quad \text{and} \quad \psi_{ub}(t)=\psi(D,t), \quad t\in[0,T] 
		\end{align}	 
	\end{subequations}
	To import PINN method for obtaining soliton solutions to the g.p. equation, we set $\psi(x,t)=u(x,t)+iv(x,t)$ where $u(x,t)$ and $v(x,t)$ represent the real and imaginary parts, respectively. Substituting this into equation~(\ref{eq:unfs_gpe}), the model equation can be expressed as follows:
	\begin{subequations}\label{eq:real_imaginary_gpe}
		\begin{align}
			f_R(x,t) &= v_{t} + \frac{1}{2}u_{xx} + 2 e^{\pm 0.02t} (u^2 + v^2) u + 0.0001 x^2 u, \label{eq:real_part_gpe} \\
			f_I(x,t) &= u_{t} - \frac{1}{2} v_{xx} - 2 e^{\pm 0.02t} (u^2 + v^2) v - 0.0001 x^2 v. \label{eq:imag_part_gpe}
		\end{align}
	\end{subequations}
	By utilizing these $f_{R}$ and $f_{I}$, we can approximate wavefunctions $\psi(x,t)$ by an artificial neural network (ANN), optimizing the training loss $L_{\text{train}}$ for the problem under consideration. The loss comprises three components, namely, loss on initial conditions ($L_I$), loss on boundary conditions ($L_B$), and loss on Residual points ($L_C$). These loss components are computed as follows:
	\begin{equation}
		L_{\text{train}} = L_{\text{IC}} + L_{\text{BC}} + L_{\text{res}}
	\end{equation}
	\begin{equation}
		L_{\text{IC}} = \frac{1}{N_I} \sum_{j=1}^{N_I} \left| \mathcal{I}[y(x_j^I, t)] \Big|_{t=0} - y^I(x_j^I) \right|^2
	\end{equation}
	\begin{equation}
		L_{\text{BC}} = \frac{1}{N_B} \sum_{j=1}^{N_B} \left| \mathcal{B}[y(x_j^B, t_j^B)] \Big|_{x^B \in \partial D} - y^B(t_j^B) \right|^2
	\end{equation}
	\begin{equation}
		L_{\text{res}} = \frac{1}{N_C} \sum_{j=1}^{N_C} f(x_j^C, t_j^C)^2
	\end{equation}
	
	In the loss function formulation of Physics-Informed Neural Networks (PINNs), $L_{\text{train}}$ represents the total training loss, which is the sum of three components: initial condition loss ($L_{\text{IC}}$), boundary condition loss ($L_{\text{BC}}$), and residual loss ($L_{\text{res}}$). The initial condition loss, $L_{\text{IC}}$, measures the discrepancy between the predicted solution $y(x,t)$ and the given initial condition $y^I(x)$ at $t = 0$, averaged over $N_I$ initial condition points. The boundary condition loss, $L_{\text{BC}}$, quantifies the error in satisfying the boundary conditions, where $\mathcal{B}[y]$ represents the boundary operator acting on the predicted solution at boundary points $x^B \in \partial D$, averaged over $N_B$ boundary points. The residual loss, $L_{\text{res}}$, enforces the underlying differential equation by minimizing the residual function $f(x,t)$, which represents the deviation of the predicted solution from satisfying the governing equation at $N_C$ collocation points. Together, these losses ensure that the neural network solution adheres to the problem's physical and mathematical constraints.		
	
	\begin{figure*}[!htbp]
		\centering
		\includegraphics[width=0.9\textwidth]{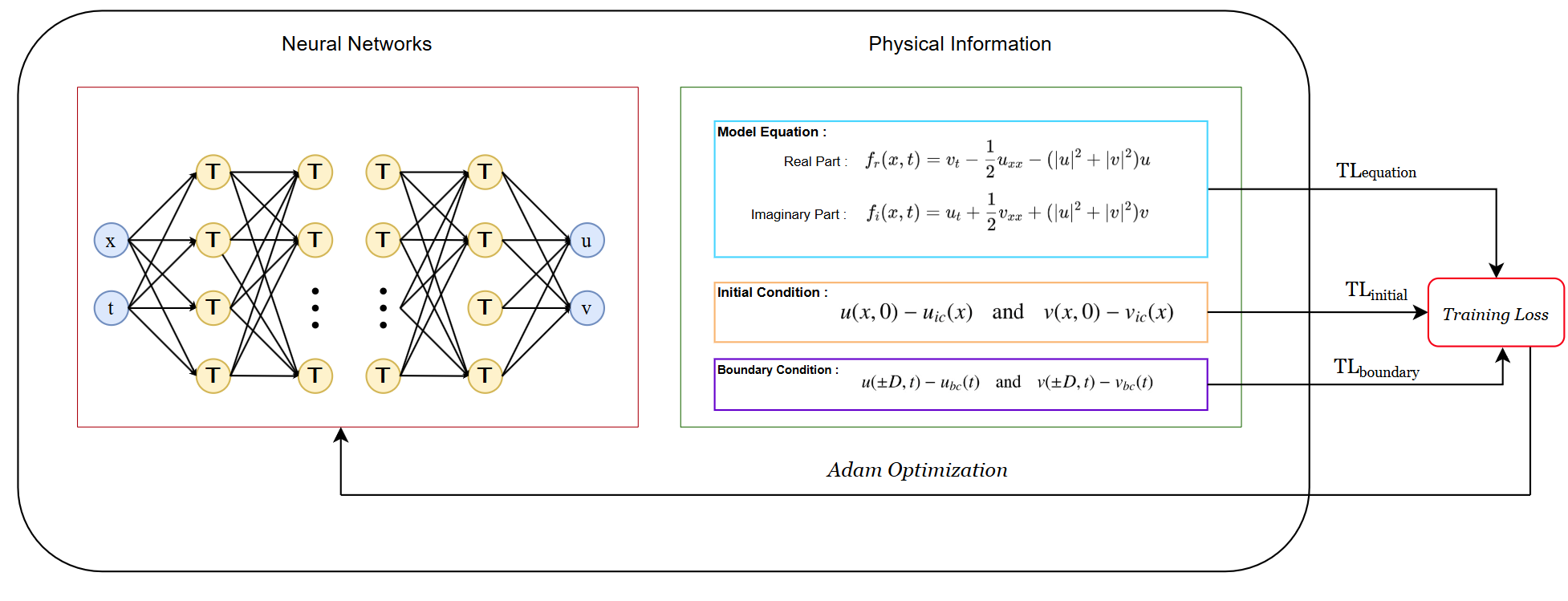}
		\caption{Schematic of the Physics-Informed Neural Network (PINN) architecture for solving the nonlinear Schrödinger equation (NLSE). The neural network takes spatial and temporal coordinates $(x, t)$ as inputs and outputs the real $(u)$ and imaginary $(v)$ parts of the solution.}
		\label{fig:pinn_nlse}
	\end{figure*}
	The schematic representation of the physics-informed neural network (PINN) is presented in Figure~\ref{fig:pinn_nlse}. It shows that there are two inputs for space and time variables, multiple hidden layers, and an output layer with neurons corresponding to the required number of wavefunction components. The values of $\psi_j$ are taken for training loss calculation. The training loss is optimized using the Adam optimizer, which is well-suited for handling non-convex optimization problems and adaptively adjusts the learning rate. These steps are repeated iteratively until the training loss is minimized. The detailed network architecture and training hyperparameters for the GPE model are summarized in Table~\ref{tab:architecture_gpe}.
	
	\begin{table}[!htbp]
		\centering
		\caption{PINN Architecture and Training Setup for GPE}
		\label{tab:architecture_gpe}
		\begin{tabular}{|l|p{4cm}|}
			\hline
			\textbf{Component} & \textbf{Details} \\ \hline
			Input Layer & 2 neurons (spatial $x$, temporal $t$) \\ \hline
			Hidden Layers & 4 layers with [50, 100, 100, 50] neurons \\ \hline
			Activation Function & $\tanh$ for all hidden layers \\ \hline
			Output Layer & 2 neurons (real-valued scalar output ($u$, $v$)) \\ \hline
			Optimizer & Adam optimizer \\ \hline
			Initial Learning Rate & 0.001 \\ \hline
			Learning Rate Scheduler & ReduceLROnPlateau (patience=500, factor=0.5) \\ \hline
			Epochs & 20,000 \\ \hline
			Loss Function & Mean Squared Error (MSE) \\ \hline
			Precision & Mixed precision training using AMP (torch.amp) \\ \hline
			Weight Initialization & PyTorch default (likely Kaiming uniform) \\ \hline
		\end{tabular}
	\end{table}
	
	\begin{itemize}
		\item \textbf{Spatial and Temporal Resolution}: 200 points in each direction (uniform grid)
		\item \textbf{Total Collocation Points}: $200 \times 200 = 40,000$
		\item \textbf{Sampling Method}: Uniform grid (no randomization or LHS)
	\end{itemize}
	
	The PINN model designed to solve the one-dimensional Gross–Pitaevskii equation employs a fully connected feedforward architecture with four hidden layers consisting of 50, 100, 100, and 50 neurons, respectively. The input layer accepts two features — the spatial ($x$) and temporal ($t$) coordinates — and the output is a single scalar function $\psi(x,t)$ representing the real-valued wavefunction amplitude.
	
	Training was performed using the Adam optimizer with an initial learning rate of 0.001, dynamically reduced using the ReduceLROnPlateau scheduler with a factor of 0.5 and a patience of 500 epochs. The total number of training epochs was 20,000. The loss function used was the mean squared error (MSE) between the predicted and exact solution values. Mixed precision training was enabled via PyTorch's automatic mixed-precision (AMP) package for computational efficiency on GPU. The domain was discretized using a uniform 2D mesh over $x \in [0, 20]$ and $t \in [0, 10]$, yielding 40,000 total collocation points.
	
	\subsection{Nonlinear Schr\"odinger Equation (NLSE)}
	
	The Nonlinear Schrödinger equation (NLSE) serves as a theoretical framework for describing optical pulse propagation in nonlinear media, Bose-Einstein condensates, water waves, and many other nonlinear phenomena~\cite{morgan1997}. Its generalized form is given by:
	\begin{equation}\label{eq:general_nlse}
		i \psi_{t} + \frac{1}{2}\psi_{xx}+ \sigma |\psi|^2 \psi=0,
	\end{equation}
	where $\sigma$ is a real constant, $\psi(x,t)$ is a complex function, and the subscripts denote partial spatial and time derivatives. In our case, we consider our system with $\sigma$ as 1, which corresponds to a focusing cubic nonlinearity in the following form.
	\begin{equation}\label{eq:model_nlse}
		i \psi_{t} + \frac{1}{2}\psi_{xx} \pm |\psi|^2 \psi=0, \quad x\in[-D,D], \quad t\in[0,T]
	\end{equation}
	The initial condition $\psi_{ic}$, lower boundary condition $\psi_{lb}$ and upper boundary condition $\psi_{ub}$ in the domain $D$ under consideration for the above model equation are given as 
	\begin{subequations}\label{eq:conditions_nlse}
		\begin{align}
			\psi_{ic}(x)&=\psi(x,0), \quad x\in[-D,D],\\
			\psi_{lb}(t)&=\psi(-D,t), \quad \text{and} \quad \psi_{ub}(t)=\psi(D,t), \quad t\in[0,T] 
		\end{align}	 
	\end{subequations}
	To import Physics Informed Neural Network for obtaining soliton solutions to the single component NLSE, we set $\psi(x,t)=u(x,t)+iv(x,t)$ where $u(x,t)$ and $v(x,t)$ represent the real and imaginary parts, respectively. Substituting this into equation~(\ref{eq:model_nlse}), the model equation can be expressed as follows:
	\begin{subequations}\label{eq:fr_fi}
		\begin{align}
			f_{r}(x,t)&=v_{t}-\frac{1}{2}u_{xx}-(|u|^2+|v|^2)u,\\
			f_{i}(x,t)&=u_{t}+\frac{1}{2}v_{xx}+(|u|^2+|v|^2)v.
		\end{align}	 
	\end{subequations}
	
	The specific architecture and parameters used for the NLSE simulation are listed in Table~\ref{tab:architecture_nlse}.
	
	\begin{table}[!htbp]
		\centering
		\caption{PINN Architecture and Training Setup for NLSE}
		\label{tab:architecture_nlse}
		\begin{tabular}{|l|p{4cm}|}
			\hline
			\textbf{Component} & \textbf{Details} \\ \hline
			Input Layer & 2 neurons ($x$, $t$) \\ \hline
			Hidden Layers & 4 layers with 128 neurons each \\ \hline
			Activation Function & $\tanh$ for all hidden layers \\ \hline
			Output Layer & 2 neurons (real and imaginary parts of $\psi$) \\ \hline
			Output Format & $\psi(x,t) = u(x,t) + i v(x,t)$ \\ \hline
			Optimizer & Adam optimizer \\ \hline
			Initial Learning Rate & 0.001 \\ \hline
			Epochs & 10,000 \\ \hline
			Loss Function & MSE of PDE residual + MSE of initial condition \\ \hline
			Initial Condition & Bright soliton: $\psi(x, 0) = \text{sech}(x)$ \\ \hline
			Training Strategy & Physics loss from NLSE + initial and boundary condition loss \\ \hline
			Precision & Single precision (32-bit float), AMP not used \\ \hline
		\end{tabular}
	\end{table}
	
	\begin{itemize}
		\item \textbf{Collocation Points:} 5,000 points and 500 boundary points in space and time
		\item \textbf{Sampling Strategy:} Uniform random sampling (not grid-based)
		\item \textbf{Residual Loss:} $\mathcal{L}_{\text{physics}} = \left| i\psi_t + \frac{1}{2} \psi_{xx} + |\psi|^2\psi \right|^2$
		\item \textbf{Initial Condition Loss:} $\mathcal{L}_{\text{IC}} = \left| \psi(x, 0) - \psi_0(x) \right|^2$
	\end{itemize}
	
	The Physics-Informed Neural Network (PINN) designed to solve the focusing Nonlinear Schrödinger Equation (NLSE) employs a fully connected feedforward architecture with four hidden layers, each containing 128 neurons and using the $\tanh$ activation function. The input layer takes two variables—space ($x$) and time ($t$)—and the output layer produces two values corresponding to the real and imaginary parts of the complex wavefunction $\psi(x,t)$.
	
	Training is performed using the Adam optimizer with an initial learning rate of 0.001 for 10,000 epochs. The total loss function consists of the sum of three terms: the residual loss enforcing the NLSE and the error in satisfying the initial and boundary conditions. The residual loss is calculated using automatic differentiation of the predicted solution with respect to space and time.

    Appropriate boundary conditions were incorporated into the loss function to ensure physical consistency at the domain edges for the NLSE. To evaluate the impact of these constraints, we conducted comparative training runs both with and without explicit boundary conditions. The resulting predicted soliton profiles were highly similar in both cases. This equivalence arises because the chosen spatial domains (e.g., $x \in [-20, 20]$) are sufficiently large relative to the localized nature of the wave structures. As the single-component NLSE admits diverse localized structures with fundamentally different asymptotic behaviors, the boundary conditions were specific to each soliton type. For the localized Bright Soliton, Dirichlet and Neumann conditions ($u=0$ and $\frac{\partial u}{\partial x}=0$) were enforced at the far spatial boundaries $x_{\min}$ and $x_{\max}$. Conversely, for the Dark Soliton, which exists on a non-vanishing continuous wave background, a Dirichlet magnitude condition ($|u|^{2}=1$) alongside a zero-gradient Neumann condition was applied. For the complex Peregrine Soliton and Akhmediev Breather, the boundaries were constrained using their respective asymptotic plane-wave backgrounds and exact spatial-temporal Dirichlet evaluations.
	
	\section{Data driven soliton: Bright, Dark, Peregrine Solitons, and Akhmediev Breathers}
	\label{sec:data}
	
	In this section, we examine the bright matter-wave soliton solution with increasing and decreasing amplitude, which serves as an analytical solution for the Gross-Pitaevskii (GP) equation~(\ref{eq:unfs_gpe}). Additionally, we explore the following localized structures: Bright Solitons, Dark Solitons, Peregrine Solitons, and Akhmediev Breathers, which serve as soliton solutions for the Nonlinear Schrödinger Equation~(\ref{eq:model_nlse}). Furthermore, we present the solution predicted by PINN Framework alongside the exact solutions.
	
	\subsection{Exact Solutions for GPE}
	
	We begin with the exact solution for the Gross-Pitaevskii (GP) equation~(\ref{eq:unfs_gpe}), as presented in~\cite{radha2007}. 
	
\begin{figure*}[!htbp]
		\centering
		\includegraphics[width=0.95\textwidth]{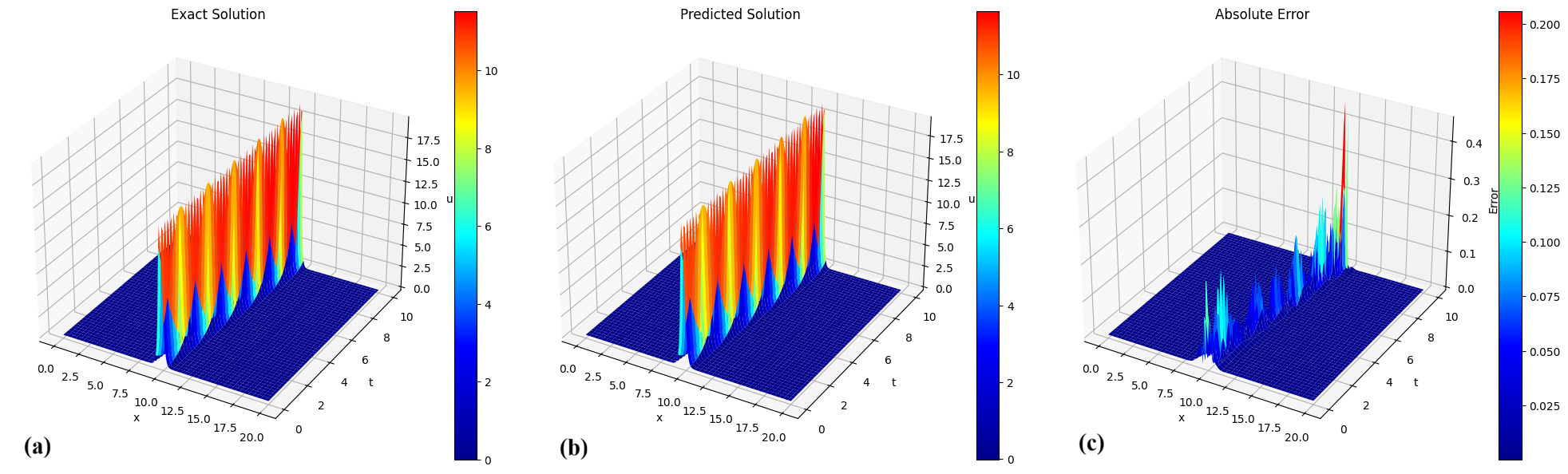}
		\caption{3D surface visualization of exact and PINN solutions for the GPE soliton solution given by \eqref{eq:exact_solution_gpe}}
		\label{fig:gpe_3d}
	\end{figure*}

	The bright matter wave solution for GPE is given in the following form
	\begin{equation}\label{eq:exact_solution_gpe}
		\psi(x,t) = 2\beta_0\exp \left( \frac{\zeta  t}{2} - \frac{i  \zeta  x^2}{4} \right)  \sech(\theta_1)\exp(i \xi_1).
	\end{equation}
	where,
	\begin{subequations}
		\begin{align}
			\xi_1 &= (2  \alpha_1 x) - \left( 4 \int (\alpha_1 - \beta_1) (\alpha_1 + \beta_1) \, dt \right) - (2  \phi_1),\notag\\
			\Theta_1 &= (2  \beta_1  x) - \left( 8 \int (\alpha_1 \beta_1) \, dt \right) + (2  \delta_1),\notag\\
			\alpha_{1}&=\alpha_{0}e^{(\zeta t)}, \quad \text{and} \quad \beta_{1}=\beta_{0}e^{(\zeta t)}.\notag
		\end{align}
	\end{subequations}  
	The equation allows for two distinct cases, depending on the choice of the parameter $\zeta$. The $\zeta$  parameter determines whether the bright soliton train exhibits a decreasing or increasing matter-wave intensity, and its interesting dynamical behaviours have been explored in he ref. \cite{radha2007}. Our goal is to check and verify whether our PINN predicted solitons exhibit the same behaviour as such in the reference  \cite{radha2007}. We set the parameters as $\alpha_0 = 0.1$, $\delta_1 = 0.5$, $\phi_1 = 0.1$ and chosen the spectral parameter which relates the external trapping, interaction strength as well as the dispersion as $\zeta$=0.02, which leads to the decreasing amplitude bright soliton profile has been shown in figure ~\ref{fig:gpe_3d} (a). Our PINN predicted soliton profile shown in ~\ref{fig:gpe_3d} (b) exactly matches the exact analytical soliton profile ~\ref{fig:gpe_3d} (a) and the respective error mismatch between the analytical and PINN predicted profile shown in ~\ref{fig:gpe_3d} (c).
    
	\subsection{Exact Solution for NLSE}
	
	The single-component more general nonlinear Schrödinger equation with positive focusing and negative defocussing nonlinearity~(\ref{eq:model_nlse}) admits various distinct types of soliton solutions. To predict soliton solutions using the PINN method, we consider the following solitons: bright solitons, dark solitons, Akhmediev breathers, and the first-order Peregrine soliton. The exact solutions for these cases are given below, which we considered from the reference ~\cite{alkhawaja2024}.
	
	\subsubsection{Bright Solitons}
   
   Considering the single-component NLS with positive nonlinearity (focusing) admits the following bright soliton solution of the following form:
	\begin{equation}
		\psi(x,t) = A_0 \sqrt{\frac{-2 a_1}{a_2}}\sech \left[ A_0 (x - x_0) \right] e^{i \left[ a_1 A_0^2 (t - t_0) + \phi_0 \right]}
	\end{equation}
	where $a_1 a_2 <0$, $A_0=1$, $a_1=0.5$, $a_2=1$, and $x_0=t_0=\phi_0=0$.
	
	\subsubsection{Dark Solitons}

	Changing the sign from positive to negative in the interaction leads to a completely different nature of the solitons. In optical language, we term it as defocusing NLS equation, which admits dark solitons. Dark solitons are dips in an otherwise continuous wave background of the following form : ~\cite{alkhawaja2024}
	\begin{equation}
		\psi(x,t) = A_0 \sqrt{\frac{-2 a_1}{a_2}} \tanh \left[ A_0 (x - x_0) \right] e^{-i \left[ 2 a_1 A_0^2 (t - t_0) + \phi_0 \right]}
	\end{equation}
	where $a_1 a_2 <0$, $A_0=1$, $a_1=0.5$, $a_2=-1$, and $x_0=t_0=\phi_0=0$.
	
	\subsubsection{Peregrine Solitons}
	
	The Peregrine soliton is a localized structure that appears transiently, characterized by its rogue wave-like behavior. It is given by:
	\begin{equation}
		\psi(x,t) = \frac{1}{\sqrt{a_2}} \left[ \frac{4 + i 8 (t - t_0)}{1 + 4 (t - t_0)^2 + \frac{2}{a_1} (x - x_0)^2} - 1 \right] e^{i \left[ t - t_0 + \phi_0 \right]}
	\end{equation}
	where $a_1=1$, $a_2=1$, and $x_0=t_0=\phi_0=0$.
	
	\subsubsection{Akhmediev Breathers}
	
	Akhmediev breathers are periodic structures that arise from the modulation instability of a continuous wave background. The solution is:
	{\small
		\begin{align}
			\psi(x,t) =& \frac{1}{\sqrt{a_2}} \left\{ 
			\frac{\kappa^2 \cosh[\delta (t - t_0)] + 2 i  \kappa  \nu  \sinh[\delta (t - t_0)]}
			{2 \cosh[\delta (t - t_0)] - 2 \nu \cos \left[ \frac{\kappa}{\sqrt{2 a_1}} (x - x_0) \right]} 
			- 1 \right\} \times \notag\\
            &e^{i \left[ t - t_0 + \phi_0 \right]}
		\end{align}
	}
	
	where $a_1 a_2 >0$, $\kappa = 2 \sqrt{1 - \nu^2}$, $\delta = \kappa \nu$, $\nu < 1$, $a_1=0.5$, $a_2=1$, $\nu=0.5$ and $x_0=t_0=\phi_0=0$.
	
	\section{Prediction of Solitons: by PINN Framework}
	
	\subsection{PINN predicted Bright soliton for GP Equation}
	
	The Gross-Pitaevskii equation (GPE) was solved using the Physics-Informed Neural Network (PINN) approach, leveraging a neural network architecture with an input layer consisting of two neurons representing spatial and temporal variables $(x,t)$, followed by four hidden layers with 100, 100, 100, and 50 neurons respectively, each employing the hyperbolic tangent ($\tanh$) activation function. The output layer consists of a two neurons representing the real and imaginary parts ($u$, $v$). The training process utilized 10,000 iterations, with a spatial domain discretized into 200 points and a temporal domain into 200 points, yielding 40,000 collocation points. Optimization was performed using the Adam optimizer with an initial learning rate of 0.001, which was adaptively reduced by a factor of 0.5 upon encountering a plateau in loss reduction, with a patience threshold of 500 epochs.
	
	The exact analytical solution of the Gross-Pitaevskii equation was derived for validation. The general form of the soliton solution for the one-dimensional GPE is given as:
	\begin{equation}
		\psi_{\text{exact}}(x,t) = A \sech(B(x - v t)) e^{i(\kappa x - \omega t)}
	\end{equation}
	where $A$ represents the amplitude, $B$ controls the width of the soliton, $v$ is the velocity, $\kappa$ is the wavenumber, and $\omega$ is the frequency term associated with the dispersion relation. The PINN-predicted solution, denoted as $\psi_{\text{PINN}}(x,t)$, was quantitatively compared to the exact solution.
	
	The absolute error between the exact and predicted solutions was computed as:
	\begin{equation}
		E_{\text{abs}}(x,t) = |\psi_{\text{PINN}}(x,t) - \psi_{\text{exact}}(x,t)|
	\end{equation}
	
	Error analysis metrics were computed to validate the accuracy of the PINN framework:
	\begin{itemize}
		\item \textbf{$L_1$ Error (MAE)}:
		\begin{equation}
			L_1 = \frac{1}{N} \sum_{i=1}^{N} | \psi_{\text{PINN}, i} - \psi_{\text{exact}, i} |
		\end{equation}
		\item \textbf{$L_2$ Error (RMSE)}:
		\begin{equation}
			L_2 = \sqrt{\frac{1}{N} \sum_{i=1}^{N} (\psi_{\text{PINN}, i} - \psi_{\text{exact}, i})^2}
		\end{equation}
		\item \textbf{Mean Error}:
		\begin{equation}
			\bar{E} = \frac{1}{N} \sum_{i=1}^{N} (\psi_{\text{PINN}, i} - \psi_{\text{exact}, i})
		\end{equation}
		\item \textbf{Maximum Absolute Error}:
		\begin{equation}
			E_{\max} = \max_{i} | \psi_{\text{PINN}, i} - \psi_{\text{exact}, i} |
		\end{equation}
	\end{itemize}
	
	The computed values for these metrics were:
	\begin{equation}
		\begin{aligned}
			L_1      &= 0.003687,  &\quad
			L_2      &= 0.017161,  \\
			\bar{E}  &= 0.000097,  &\quad
			E_{\max} &= 0.455342.
		\end{aligned}
	\end{equation}

	These results indicate a high degree of accuracy in the predicted solution, with the mean error approaching zero, signifying the absence of systematic bias. The maximum absolute error, though relatively higher, was localized to specific regions, likely corresponding to sharp transitions where the $\tanh$ activation function struggles to approximate steep gradients effectively.
	
\begin{figure*}[!htbp]
		\centering
		\includegraphics[width=0.9\textwidth]{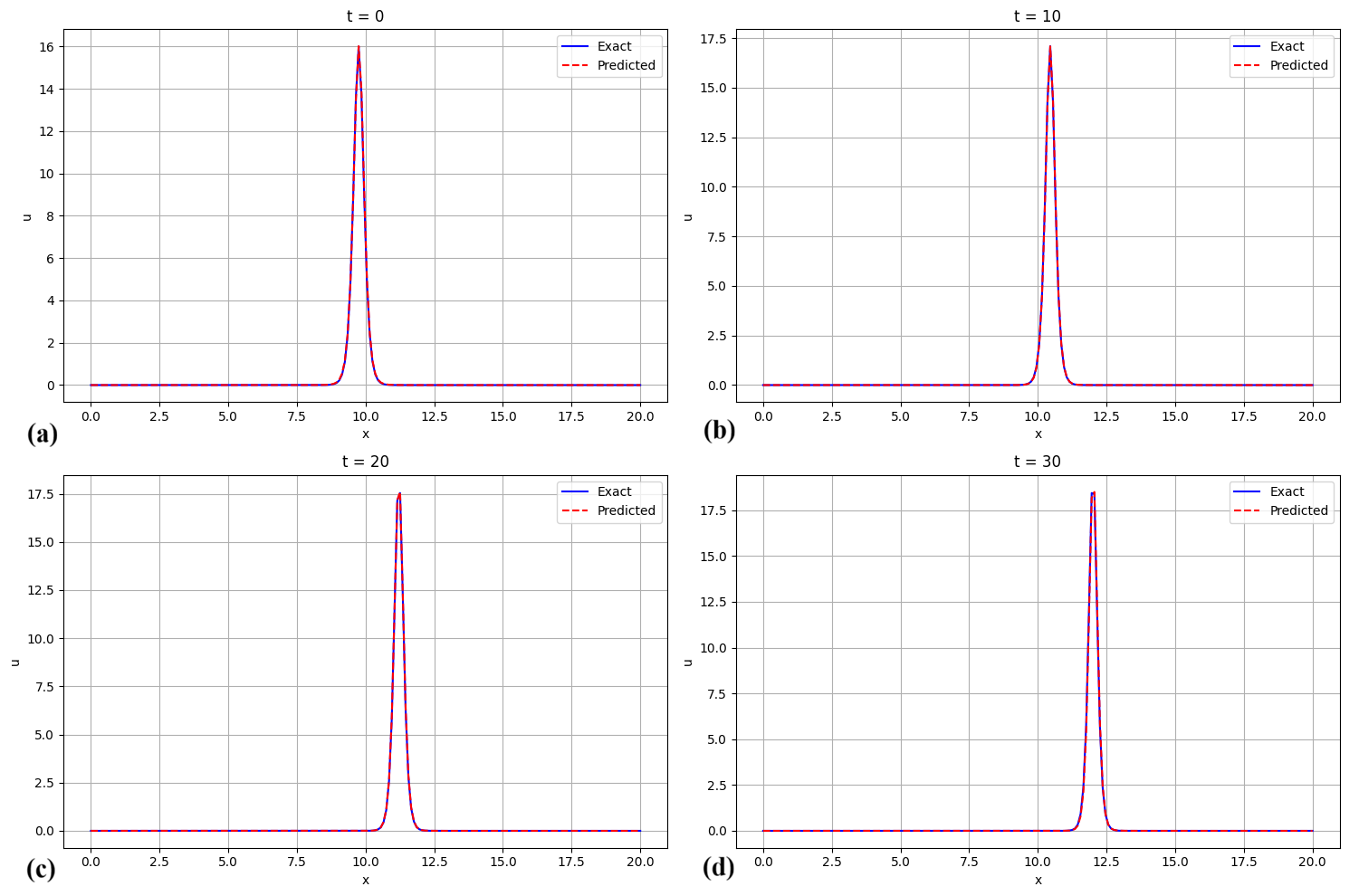}
		\caption{Exact vs. PINN-predicted $u(x,t)$ at $t = 0$, 10, 20, and 30, demonstrating precise soliton dynamics captured during the examination of GP Equation.}
		\label{fig:training_plot}
	\end{figure*}

	The loss landscape during training, as depicted in Figure~\ref{fig:training_plot}, demonstrates stable convergence towards a minimum, confirming the effectiveness of the training strategy. Figure~\ref{fig:training_plot}, (a-d) illustrates the agreement between the convergence between exact and predicted for different time intervals. The $\ell_2$-norm error between the exact and predicted solutions remained within an acceptable tolerance, particularly in regions where the solution exhibited smooth variations. Slight deviations were observed in regions of steep gradients, reinforcing known limitations of the activation function in capturing rapid transitions.
	
\begin{figure*}[!htbp]
		\centering
		\includegraphics[width=0.8\textwidth]{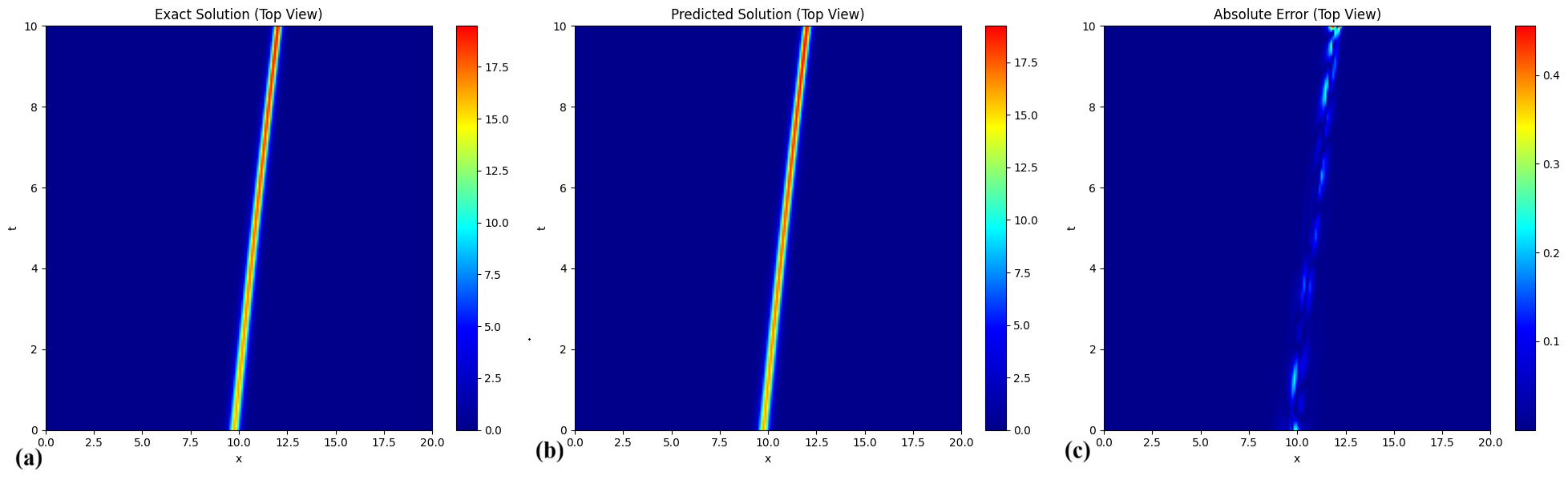}
		\caption{Top-view comparison of exact and PINN-predicted solutions for the GPE soliton solution. The absolute error remains below 0.5 across the domain, validating the prediction fidelity.}
		\label{fig:gpe_topview}
	\end{figure*}

	The results demonstrate that the Physics-Informed Neural Network (PINN) framework effectively captures the intricate dynamics of the Gross-Pitaevskii (GP) equation as shown in 2D and 3D Figures~\ref{fig:gpe_topview} and~\ref{fig:gpe_3d} respectively. Parameters: soliton speed $c = 4.0$, domain $x \in [0, 20]$, $t \in [0, 30]$, initial condition $u(x,0) = \frac{c}{2} \, \sech^2 \left(\frac{\sqrt{c}}{2}(x - x_0)\right)$ with $x_0 = 10$, PINN architecture: 3 hidden layers $\times$ 50 neurons, tanh activation. Maximum absolute error $\approx 0.45$ and accurately resolving the interplay between dispersion and nonlinearity. The trained model successfully reproduces soliton interactions, wavefunction narrowing, and amplitude enhancement over time, aligning well with analytical expectations. The continuous, differentiable representation offered by PINNs ensures a smooth approximation of the solution while maintaining physical consistency. Additionally, the model exhibits strong generalization capabilities, preserving the fundamental characteristics of the GP equation without relying on explicit numerical discretization schemes.
	
\section{PINN Predicted Family of solitons for NLSE}
\subsection{Bright soliton}
\begin{figure*}
\centering
\includegraphics[width=0.8\textwidth]{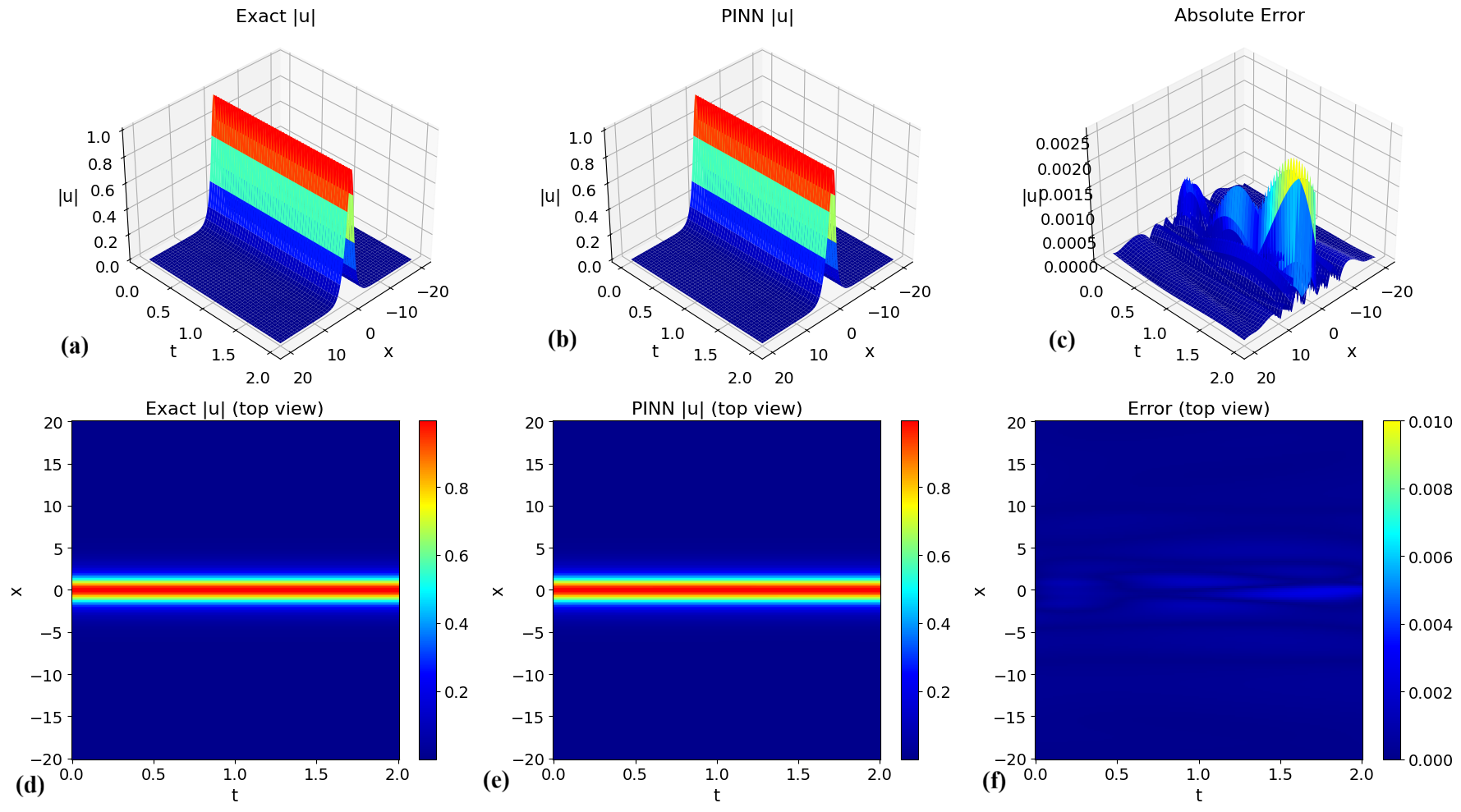}
\caption{Bright soliton propagation results.
(\textbf{a}) 3D surface visualizations of the exact analytical solution, (\textbf{b}) the PINN-predicted solution, and (\textbf{c}) the resulting absolute error, demonstrating high-fidelity reconstruction of the stationary pulse.(\textbf{d}) 2D heatmap projections (top view) for the exact solution (\textbf{a}),  (\textbf{e}) PINN prediction, and error (\textbf{f}). The framework successfully maintains the localized soliton shape with a maximum absolute error of approximately 0.0048.}
\label{fig:bright_soliton}
\end{figure*}
	
	In this section, we present the results obtained by solving the Nonlinear Schrödinger Equation (NLSE) for the bright soliton case using the Physics-Informed Neural Network (PINN) approach. The bright soliton solution is analytically expressed as
	\begin{equation}
		\psi(x,t) = A \sech \left( A (x - v t) \right) e^{i (v x - (\frac{v^2}{2} - A^2) t)},
	\end{equation}
	where $A$ represents the amplitude, and $v$ denotes the velocity of the soliton. The PINN framework was trained 5,000 collocation points and 500 boundary points. A neural network with four hidden layers, each containing 128 neurons, was employed, using the hyperbolic tangent activation function.
	
	The comparison between the exact analytical solution and the PINN-predicted solution is shown in Figure~\ref{fig:bright_soliton}. The predicted soliton profile maintains the expected localized nature, and the absolute error between the predicted and exact solutions remains minimal. Parametric choices: domain $x \in [-20, 20]$, $t \in [0, 2]$; soliton parameters $A = 1.0$, $B = 1.0$, $v = 0$. The soliton preserves its shape over time, indicating that the PINN approach effectively captures the fundamental characteristics of the bright soliton solution.
	
	\subsection{Dark Soliton}
	\begin{figure*}[!htbp]
		\centering
		\includegraphics[width=0.8\textwidth]{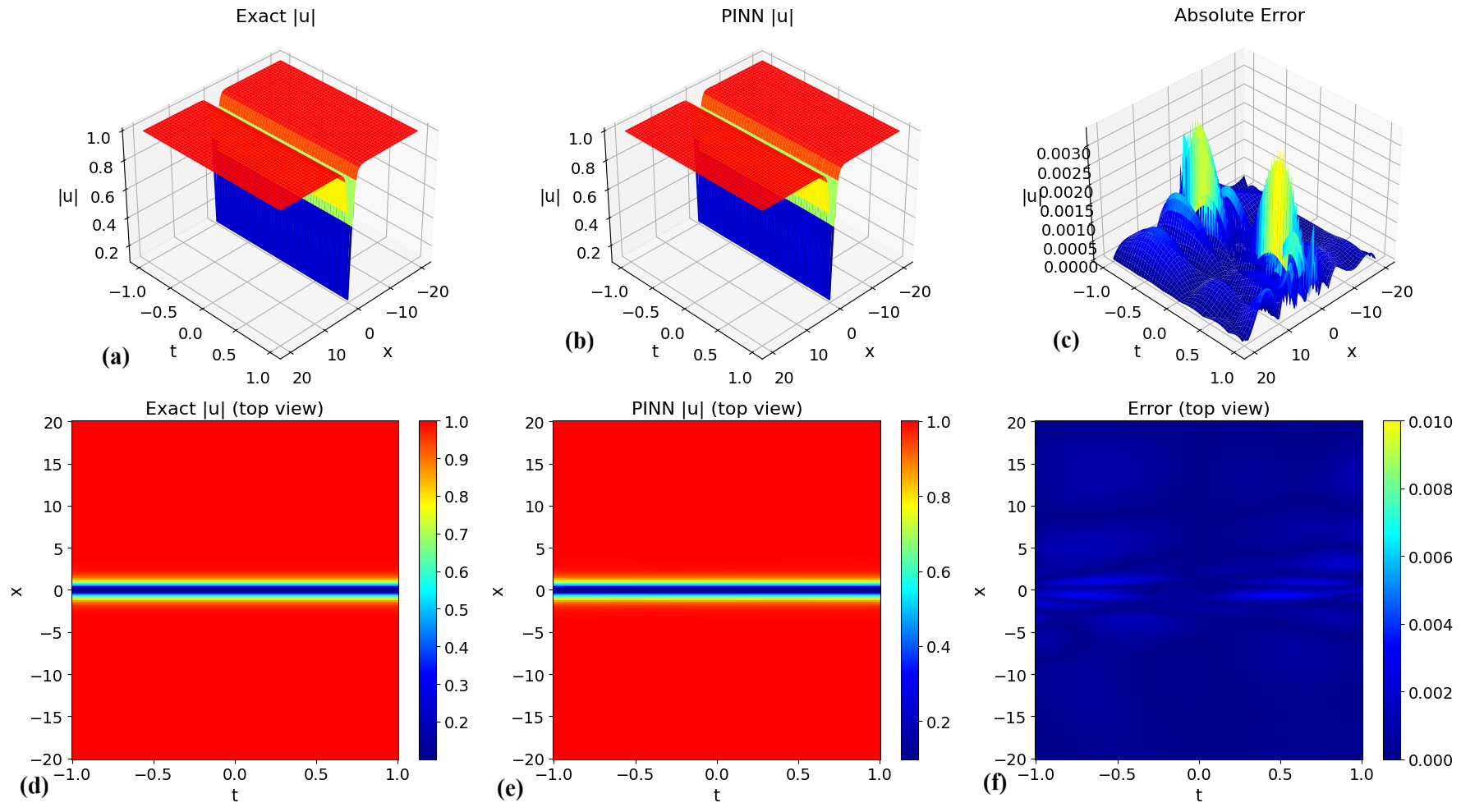}
	\caption{Dark soliton propagation results.
    (\textbf{a}) 3D surface visualizations of the exact analytical solution, the PINN-predicted solution (\textbf{b}), and (\textbf{c}) the resulting absolute error, demonstrating high-fidelity reconstruction of the stationary pulse. (\textbf{d}) 2D heatmap projections (top view) for the exact solution, PINN prediction (\textbf{e}), and error (\textbf{f}). The framework successfully maintains the localized soliton shape with a maximum absolute error of approximately 0.0033.}
	\label{fig:dark_soliton}
\end{figure*}
	The exact dark soliton solution is given by
	\begin{equation}
		\psi(x,t) = A \tanh \left( A (x - v t) \right) e^{i (v x - (\frac{v^2}{2} - A^2) t)},
	\end{equation}
	where $A$ represents the background amplitude, and $v$ is the soliton velocity. Similar to the bright soliton case, the PINN was trained with 5,000 collocation points and 500 boundary points, employing the same neural network architecture and optimization strategy and parametric choices: domain $x \in [-20, 20]$, $t \in \textcolor{blue}{[-1, 1]}$; soliton parameters $A = 1.0$, $B = 1.0$, $v = 0$.
	
	The comparison between the analytical and PINN-predicted solutions is presented in Figure~\ref{fig:dark_soliton}. The results indicate that the PINN framework accurately reproduces the dark soliton's characteristic dip in the amplitude profile while maintaining the expected phase structure.

	\subsection{Peregrine Soliton}
	\begin{figure*}[!htbp]
		\centering
		\includegraphics[width=0.8\textwidth]{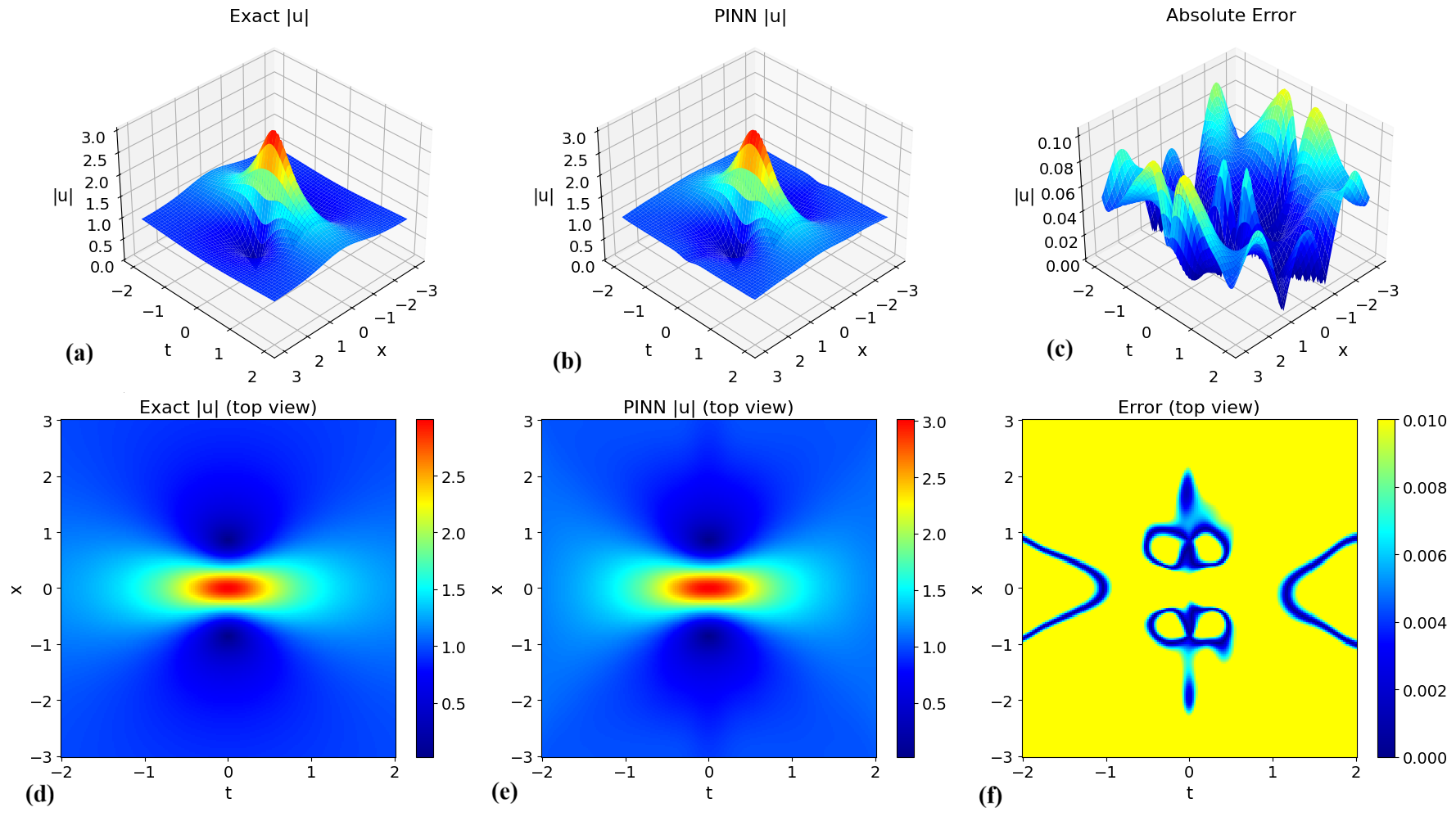}
	\caption{Peregrine soliton propagation results.(\textbf{a}) 3D surface visualizations of the exact analytical solution, the PINN-predicted solution (\textbf{b}), and (\textbf{c}) the resulting absolute error, demonstrating high-fidelity reconstruction of the stationary pulse. (\textbf{d}) 2D heatmap projections (top view) for the exact solution, PINN prediction (\textbf{e}), and error (\textbf{f}).
    The framework successfully maintains the localized soliton shape with a maximum absolute error of approximately 0.1043.}
	\label{fig:peregrine_soliton}
\end{figure*}
	
	The Peregrine soliton solution of the Nonlinear Schrödinger Equation (NLSE) was approximated using a Physics-Informed Neural Network (PINN). The computational domain was set over the spatial and temporal range $x \in \textcolor{blue}{[-3,3]}$ and $t \in \textcolor{blue}{[-2,2]}$ (Figure~\ref{fig:peregrine_soliton}).
	
	The neural network architecture consisted of an input layer with two neurons corresponding to $x$ and $t$, followed by four hidden layers, each containing \textcolor{blue}{128} neurons with a hyperbolic tangent (\texttt{tanh}) activation function, and an output layer representing $\psi(x,t)$. Training was performed using the Adam optimizer with an initial learning rate of $0.001$ for $\textcolor{blue}{10,000}$ epochs, utilizing $\textcolor{blue}{5,000}$ collocation points and \textcolor{blue}{500 boundary points} in total.
	
	The accuracy of the PINN model was assessed by computing the absolute error, defined as
	\begin{equation}
		E(x,t) = |\psi_{\text{exact}}(x,t) - \psi_{\text{PINN}}(x,t)|.
	\end{equation}
	
	The obtained error statistics are: maximum error of $0.104327$, mean error of $0.041026$, and root mean squared error (RMSE) of $0.047667$. Parametric choices: $x \in [-3, 3]$, $t \in [-2, 2]$.

	\subsection{Akhmediev breather}
	\begin{figure*}[!htbp]
	\centering
	\includegraphics[width=0.8\textwidth]{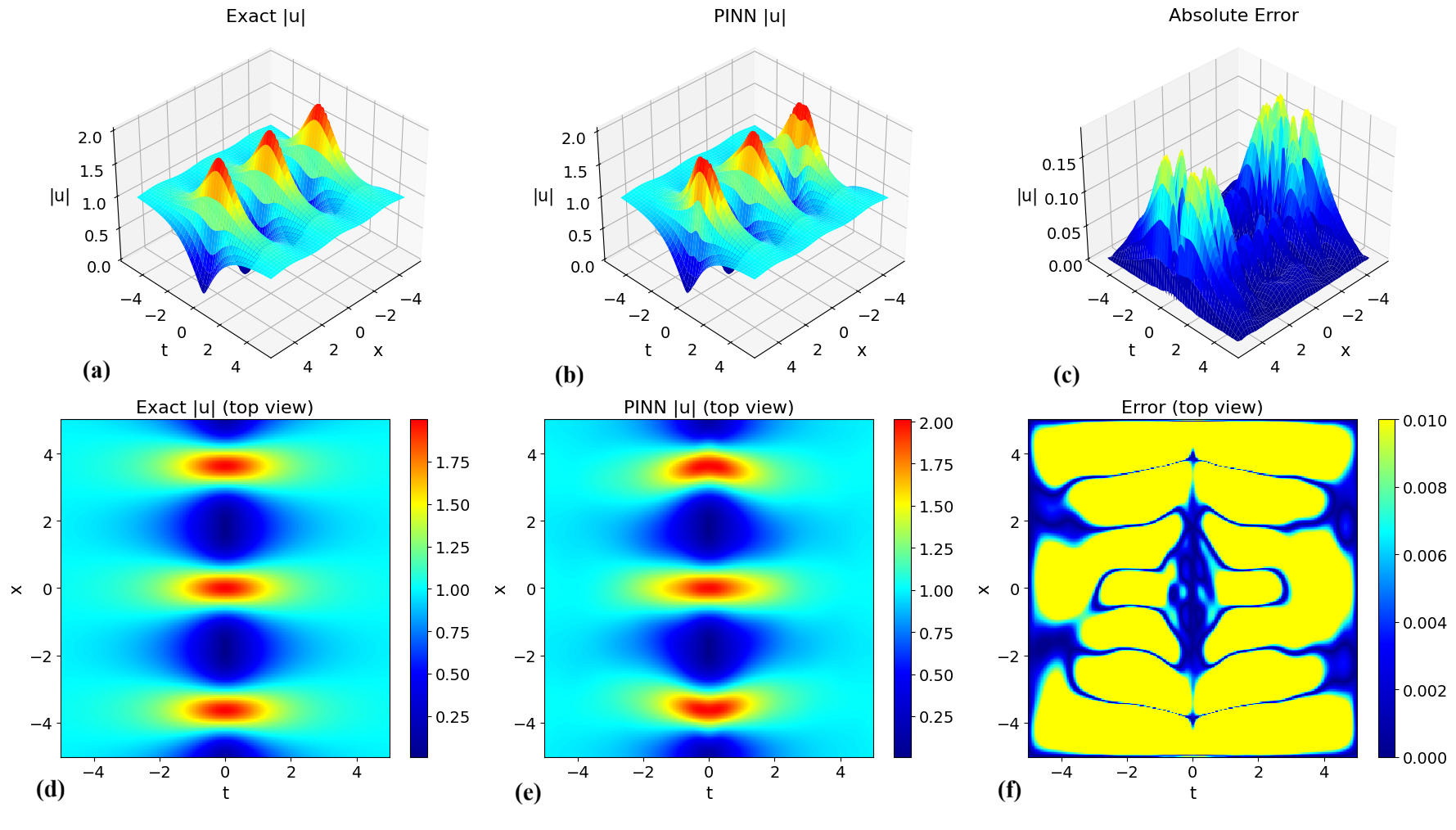}
	\caption{Akhmediev breather propagation results. (\textbf{a}) 3D surface visualizations of the exact analytical solution, the PINN-predicted solution (\textbf{b}), and (\textbf{c}) the resulting absolute error, demonstrating high-fidelity reconstruction of the stationary pulse. (\textbf{d}) 2D heatmap projections (top view) for the exact solution, PINN prediction (\textbf{e}), and error (\textbf{f}). The framework successfully maintains the localized soliton shape with a maximum absolute error of approximately 0.187.}
	\label{fig:akhmediev_breather}
\end{figure*}
	The Akhmediev breather solution of the Nonlinear Schrödinger Equation (NLSE) was approximated using a Physics-Informed Neural Network (PINN) within the computational domain $x \in \textcolor{blue}{[-5,5]}$ and $t \in \textcolor{blue}{[-5,5]}$. The neural network model was trained for $\textcolor{blue}{10,000}$ epochs using the Adam optimizer with an initial learning rate of $0.001$, utilizing $\textcolor{blue}{5000}$ collocation points and \textcolor{blue}{500 boundary points} to enforce the governing equation and boundary conditions. The absolute error between the exact and predicted solutions was computed as:
	\begin{equation}
		E(x,t) = |\psi_{\text{exact}}(x,t) - \psi_{\text{PINN}}(x,t)|.
	\end{equation}
	
	The error statistics indicate a maximum error of $0.187344$, a mean error of $0.038786$, and a root mean squared error (RMSE) of $0.057770$. The visualization of the PINN prediction, presented alongside the exact solution and the corresponding error distribution, reveals that the model successfully captures the periodic and localized structure of the Akhmediev Breather (Figure~\ref{fig:akhmediev_breather}). Parametric choices: $x \in \textcolor{blue}{[-4, 4]}$, $t \in \textcolor{blue}{[-4, 4]}$.

	\section{Summary and Future direction}
	\label{sec:summary}
	
	To assess the accuracy and generalization capability of the proposed PINN framework, we employed multiple error metrics, including mean absolute error ($L_1$), root mean squared error ($L_2$), mean error, and maximum absolute error. These were computed over the entire domain by comparing the predicted and exact analytical solutions. The dataset for each model was carefully selected based on the nature of the soliton solution: for the GPE, a structured grid of 40,000 collocation points was used, while for the NLSE, \textcolor{blue}{5,000} collocation points were used. The dataset for each model was carefully selected based on the nature of the soliton solution and an analysis of sampling sensitivity. For the GPE, tracking the external harmonic trapping potential over a prolonged temporal domain required a structured grid, naturally yielding 40,000 collocation points. Conversely, the NLSE models utilized random uniform sampling \cite{raissi2019physics}. Empirical convergence testing on the NLSE revealed that while a lower density occasionally introduced variance near steep wave gradients, scaling to 5,000 collocation points and 500 boundary points stabilized the error metrics \cite{wu2023comprehensive}. This demonstrated that the PINN framework's accuracy is sensitive to under-sampling the nonlinear interaction regions, but robust to the specific sampling strategy provided the empirical density threshold is met.

    Future work will incorporate adaptive stopping based on validation loss curves and benchmark comparisons against standard PINNs to better quantify the gains introduced by our PINN framework enhancements.

    Additionally, while the current implementation of the PINN framework demonstrates high accuracy in 1D single-component systems, scaling the framework to higher-dimensional PDEs or multi-component coupled systems presents significant challenges \cite{li2020fourier}. Specifically, standard activation functions such as $\tanh$ suffer from spectral bias, inherently favoring low-frequency mappings and struggling to accurately resolve the steep spatial gradients often found at the edges of localized soliton peaks \cite{wang2021eigenvalue}. Apart from this, multi-component systems often exhibit multi-scale spatio-temporal dynamics, resulting in stiff equations that complicate the optimization process \cite{wang2022and}. Enforcing boundary conditions in higher dimensions requires extensive sampling across complex geometric surfaces, which significantly increases computational costs, and coupled equations generate multiple competing loss terms, necessitating advanced adaptive weighting techniques to ensure the network balances training across all components effectively \cite{jagtap2020adaptive}.

    To overcome these obstacles and improve performance near steep gradients, future frameworks will be incorporated with alternative activation functions. For instance, locally adaptive activation functions (LAAFs) introduce trainable scaling parameters that dynamically adjust the activation slope, allowing the network to capture sharp transitions without requiring deeper architectures \cite{jagtap2020adaptive}. Similarly, spectral or periodic activation functions (such as SIREN) can be utilized to better represent high-frequency solution components by maintaining non-vanishing derivatives \cite{sitzmann2020implicit}. Beyond activation functions, implementing hybrid architectures such as Fourier Neural Operators \cite{li2020fourier} or DeepONets \cite{lu2021learning} can improve generalization. Techniques like domain decomposition, adaptive mesh refinement, and residual-based collocation sampling may also be integrated to enhance accuracy and efficiency, particularly for extending applicability to complex quantum systems.
	
	\section{Declaration of Competing Interest}
	The authors declare that they have no known competing financial interests or personal relationships that could have appeared to influence the work reported in this paper.
	
	\section{Acknowledgements}
	PSV wishes to express his deepest gratitude to the Management of PSG College of Arts and Science for their moral support and encouragement throughout the tenure of this project.
	
	\section{Author contribution statement}
	\textbf{P.S.Vinayagam:} Visualization, Validation, Supervision, Conceptualization, Writing – original draft, Writing – review \& editing. \textbf{Sai Jeevanth:} Conceptualization, Software, Methodology, Investigation, Writing – review original draft. \textbf{D. Aravindha Krishnan:} Software, Methodology, Investigation and Writing – review original draft. \textbf{Nithish kathiravan:} Software, Methodology, Investigation, Conceptualization and Validation.
	
	\section{Data availability statement}
	The data used to describe the content of the article is available within the article.
 
\end{document}